\documentclass[pra,twocolumn,superscriptaddress]{revtex4}
\usepackage{amsbsy}
\usepackage{latexsym,epsfig,graphicx}
\usepackage{dcolumn}
\usepackage{graphicx}
\usepackage{subfigure}
\usepackage{comment}
\usepackage{color}
\usepackage{bm}
\usepackage{mathrsfs}
\usepackage{amsfonts}
\usepackage{amsmath}
\usepackage{color}
\usepackage{amssymb}
\usepackage{xspace}
\usepackage{epstopdf}
\usepackage{tabularx}
\usepackage{longtable}
\usepackage[colorlinks=true, letterpaper=true, pdfstartview=FitV, linkcolor=blue, citecolor=blue, urlcolor=blue]{hyperref}
\usepackage[normalem]{ulem}

\pdfoutput=1
\input{tcilatex}
\begin{document}

\title{Bogoliubov Corner Excitations in Conventional $s$-Wave Superfluids}
\author{Wei Tu}
\affiliation{School of Sciences, Xi'an Technological University, Xi'an 710032, China}
\author{Ya-Jie Wu}
\thanks{wuyajie@xatu.edu.cn}
\affiliation{School of Sciences, Xi'an Technological University, Xi'an 710032, China}
\author{Ning Li}
\affiliation{School of Sciences, Xi'an Technological University, Xi'an 710032, China}
\author{Miaodi Guo}
\affiliation{School of Sciences, Xi'an Technological University, Xi'an 710032, China}
\author{Junpeng Hou}
\affiliation{Pinterest Inc., San Francisco, California 94103, USA}

\begin{abstract}
Higher-order topological superconductors and superfluids have triggered a
great deal of interest in recent years. While Majorana corner or hinge
states have been studied intensively, whether superconductors and
superfluids, being topological or trivial, host higher-order topological Bogoliubov excitations remains elusive. In this work, we propose that Bogoliubov corner
excitations can be driven from a trivial conventional $s$-wave superfluid
through mirror-symmetric local potentials. The topological Bogoliubov
excited modes originate from the nontrivial Bogoliubov excitation bands.
These modes are protected by mirror symmetry and robust against
mirror-symmetric perturbations as long as the Bogoliubov energy gap remains
open. Our work provides new insight into higher-order topological excitation
states in superfluids and superconductors.
\end{abstract}

\maketitle


\section{Introduction and Motivation}

Topological phases have ignited intensive research interests in the past two
decades. Intrinsic topological states with $n$-th order in $d$ dimension exhibit $%
d-n$ dimensional gapless boundary states. Due to the bulk-boundary
correspondence, the nontrivial bulk topology for the higher-order
topological states ($n>1$) is different from the conventional ($n=1$)
topological states \cite{Benalcazar2017,Langbehn2017,Song2017}. The
celebrated tenfold way can characterize the first-order topological
insulators and superconductors in a unified way in terms of three nonspatial
symmetries, i.e., time-reversal, particle-hole, and chiral symmetries \cite%
{Schnyder2008,Kitaev2009,Ryu2010}. However, higher-order topological
states are usually related to crystalline symmetries, and the comprehensive
topological classifications have been made recently with point group
symmetries \cite{Khalaf2018,Miert2018,Cornfeld2019,Geier2020,Cornfeld2021}.

Topological superconductors and superfluids as one type of nontrivial
topological states have attracted great attention due to nontrivial
non-Abelian Majorana modes and potential applications in topological quantum
computing. In recent years, higher-order topological superconductors and
superfluids are proposed in various platforms, such as superconductor-topological insulator
heterostructures \cite{ZYan2018,QWang2018,Hsu2018,tliu2018,Pan2019,XZhu2019,Peng2019,YWu2020,Laubscher2020}, iron-based
superconductors \cite{RZhang2019,XZhang2019,Gray2019,XiWu2020,Kheirkhah2021,Qin2022}, $\pi$-Joesphson junctions \cite%
{YVolpez2019,SBZhang2020}, ultracold atomic systems \cite{CZeng2019,Huang2019,Zhigang2019,kheir2020,YBWu2020,YJWuJ2020,YBWub2021,YBWu2023} and (twisted) bilayer
graphene \cite{Laubscherb2020,AChew2023}. Majorana (Kramers) corner or hinge modes
naturally arise as the exhibitions of non-trivial higher-order topology.
However, while the topological property of ground states for
higher-order topological superconductors has been studied intensively, the
topology for the excitation bands of the superconductors remains elusive.

In this paper, we show that Bogoliubov corner excited modes could emerge
in a conventional $s$-wave superfluid on a honeycomb lattice with the mirror-symmetric onsite potential. To gain more intuitive insight, we
first showcase an $s$-wave superconductor on one-dimensional (1D) lattice
with inversion-symmetric potential hosting topologically nontrivial edge
excitation modes, despite the ground state for the superconductor is in a trivial phase. These topological modes could extend to a two-dimensional (2D)
square lattice with a defect chain. Furthermore, the
higher-order topological Bogoliubov corner excitation modes are present in an $s$%
-wave superfluid on a 2D honeycomb lattice. These Bogoliubov
excitation modes are protected by the nontrivial topology for the
Bogoliubov excitation bands, and robust against mirror-symmetric perturbations.

The remainder of this paper is organized as follows. In Sec. \ref{s1d}, we
take a simple 1D $s$-wave superconductor as an example to show that the
inversion-symmetric onsite potential could induce localized edge excitation modes.
The topological origin of edge modes is explored and demonstrated. In Sec. \ref{s2ds}, we
consider a defect chain on a 2D square lattice exhibiting robust edge
modes. In Sec. \ref{s2h}, we consider an $s$-wave
superfluid on a honeycomb lattice with the mirror-symmetric potential, and show
the Bogliubov corner excitation modes. In Sec. \ref{con}, we discuss relevant topics including experiment realizations, and draw a conclusion.

\section{Bogliubov edge excitations in 1D $S$-wave superconductors}

\label{s1d}

\begin{figure}[tbp]
\centering\includegraphics[width=0.48\textwidth]{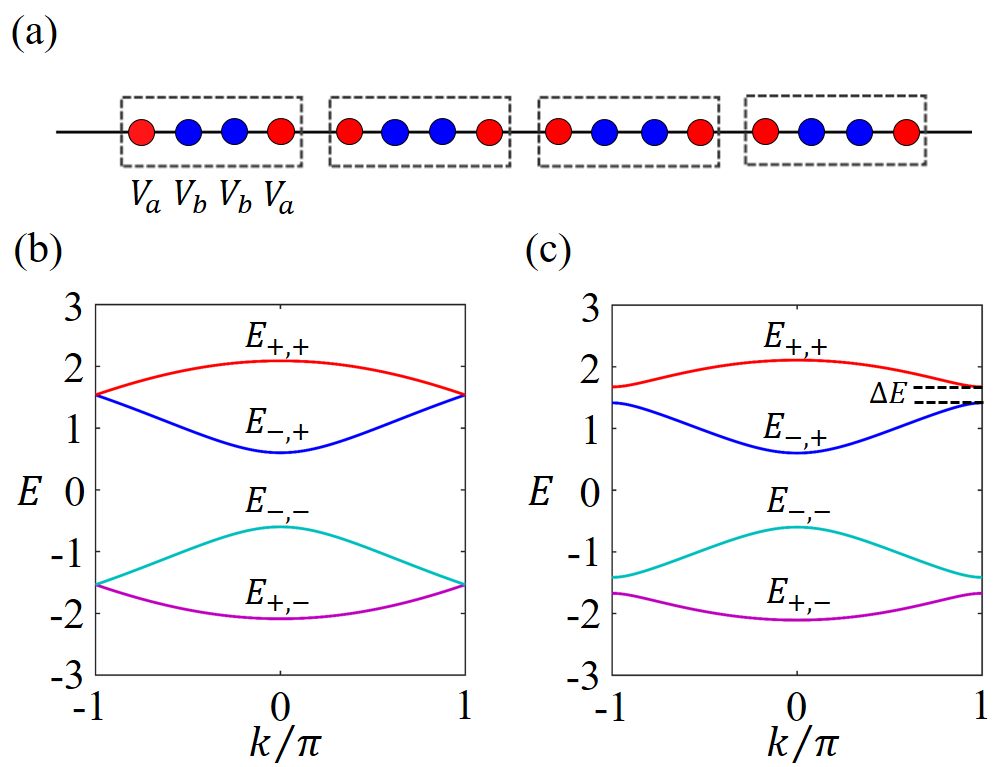}
\caption{(a) Illustration of a 1D lattice with inversion-symmetric onsite
potentials $V_{a}$ and $V_{b}$. Each unit cell consists of four sublattice
sites indexed by 1 to 4 (from left to right). (b) and (c) Energy spectra for 1D superconductor with
onsite potentials $V_{a}=-V_{b}=0$ and $V_{a}=-V_{b}=0.2$, respectively. The
superconductor order parameter is set to be $\Delta _{0}=0.6$ in (b) and
(c). Common parameter is set to be $t=1$.}
\label{1D}
\end{figure}

We first consider a simple model, i.e., 1D $s$-wave superconductor, to
present topologically protected Bogoliubov edge excitations emerging in the
presence of inversion-symmetric potentials, as shown in Fig. \ref{1D}(a). Its
physics is described by the Hamiltonian $\hat{H}=\hat{H}_{0}+\hat{H}_{V}$.
The first part reads
\begin{equation}
\hat{H}_{0}=t\sum_{\left\langle i,j\right\rangle ,\sigma }\hat{c}_{i,\sigma
}^{\dagger }\hat{c}_{j,\sigma }+\sum_{i}\left( \Delta _{0}\hat{c}%
_{i,\uparrow }^{\dagger }\hat{c}_{i,\downarrow }^{\dagger }+h.c.\right) ,
\end{equation}%
where $t$ denotes the tunneling strength between nearest neighbor sites, $%
\left\langle ...\right\rangle $ represents the summation over all
nearest-neighbor sites, $\sigma =\left( \uparrow ,\downarrow \right) $ is
the spin index, and $\Delta _{0}$ is an $s$-wave superconductor order
parameter. The second term $\hat{H}_{V}=\sum_{i}V_{i}\hat{c}_{i,\sigma }^{\dagger }\hat{%
c}_{i,\sigma }$ describes onsite potentials with inversion
symmetry. In the following, we consider each unit cell
consisting of four sublattices with the potentials $V_{i}=V_{a}$ if $\func{mod}%
\left( i,4\right) =0$ or $1$ and $V_{i}=V_{b}$ if $\func{mod}\left(
i,4\right) =2$ or $3$. For simplicity, we set $V_{a}=-V_{b}=V$ throughout
the paper if not otherwise specified. We set nearest-neighbor hopping as the unit of energy.

Through the Fourier transformation, the Hamiltonian $\hat{H}$ for the system with periodic boundary conditions
can be written as $\hat{H%
}=\sum_{\mathbf{k}}\hat{C}_{\mathbf{k}}^{\dag }H\left( \mathbf{k}\right)
\hat{C}_{\mathbf{k}}$ with%
\begin{eqnarray}
H\left( k\right) &=&ts_{x}s_{0}\sigma _{0}\tau _{z}+\xi _{+}s_{x}s_{x}\sigma
_{0}\tau _{z}+\xi _{-}s_{y}s_{y}\sigma _{0}\tau _{z}  \notag \\
&&+\xi _{0}\left( s_{x}s_{y}\sigma _{0}\tau _{z}+s_{y}s_{x}\sigma _{0}\tau
_{z}\right) -\Delta _{0}s_{0}s_{0}\sigma _{y}\tau _{y}  \notag \\
&&+Vs_{z}s_{z}\sigma _{0}\tau _{z}
\end{eqnarray}
under the basis $\hat{C}_{\mathbf{k}}^{\dagger }=\left( \hat{\psi}_{\mathbf{k%
}\uparrow }^{\dag },\hat{\psi}_{\mathbf{k}\downarrow }^{\dagger },\hat{\psi}%
_{-\mathbf{k}\uparrow },\hat{\psi}_{-\mathbf{k}\downarrow }\right) $, where  $\hat{\psi}_{\mathbf{k}\sigma }^{\dag }=$ $\left( \hat{c}_{1,%
\mathbf{k},\sigma }^{\dagger },\hat{c}_{2,\mathbf{k},\sigma }^{\dagger },%
\hat{c}_{3,\mathbf{k},\sigma }^{\dagger },\hat{c}_{4,\mathbf{k},\sigma
}^{\dagger }\right) $, $\hat{\psi}_{-\mathbf{k}\sigma }=$ $\left( \hat{c}%
_{1,-\mathbf{k},\sigma },\hat{c}_{2,-\mathbf{k},\sigma },\hat{c}_{3,-\mathbf{%
k},\sigma },\hat{c}_{4,-\mathbf{k},\sigma }\right) $, the quantity $%
\xi _{\pm }=\frac{t}{2}\left( 1\pm \cos k\right) $, and $\xi _{0}=\frac{t}{2}%
\sin k$. The Pauli matrices $\boldsymbol{s}$, $%
\boldsymbol{\sigma }$, $\boldsymbol{\tau }$ act on sublattice space, spin space and
particle-hole space, respectively, while the nought subscripts represent identity matrices.

The system preserves time-reversal ($\mathcal{T}$), particle-hole ($\mathcal{%
P}$) and chiral symmetries ($\mathcal{C}$). The energy spectra for the system are
given by
\begin{eqnarray}
E_{\pm ,+} &=&\pm \sqrt{\Delta _{0}^{2}+2t^{2}+V^{2}+2t\sqrt{t^{2}\cos ^{2}%
\frac{k}{2}+V^{2}}}, \\
E_{\pm ,-} &=&\pm \sqrt{\Delta _{0}^{2}+2t^{2}+V^{2}-2t\sqrt{t^{2}\cos ^{2}%
\frac{k}{2}+V^{2}}}.
\end{eqnarray}

Each energy level is four-fold degenerate. The energy gap for the two
excitation bands $E_{+,+}$ and $E_{-,+}$ at high-symmetry momentum point $%
k=\pi $ is $\Delta E=\sqrt{\Delta _{0}^{2}+t^{2}+\left( t+V\right) ^{2}}-%
\sqrt{\Delta _{0}^{2}+t^{2}+\left( t-V\right) ^{2}}$. In the absence of
inversion-symmetric potentials, namely $V=0$, the two Bogoliubov excitation
bands are degenerate at $k=\pi $ with $\Delta E=0$, as
illustrated in Fig. \ref{1D}(b). When $V\neq 0$, an energy
gap $\Delta E\neq 0$ is opened, as shown in Fig. \ref{1D}(c). Therefore, the introduction of
inversion-symmetric potentials opens the gap for excitation bands, which implies a topological phase transition as discussed in the following.
\begin{figure}[tbp]
\centering\includegraphics[width=0.46\textwidth]{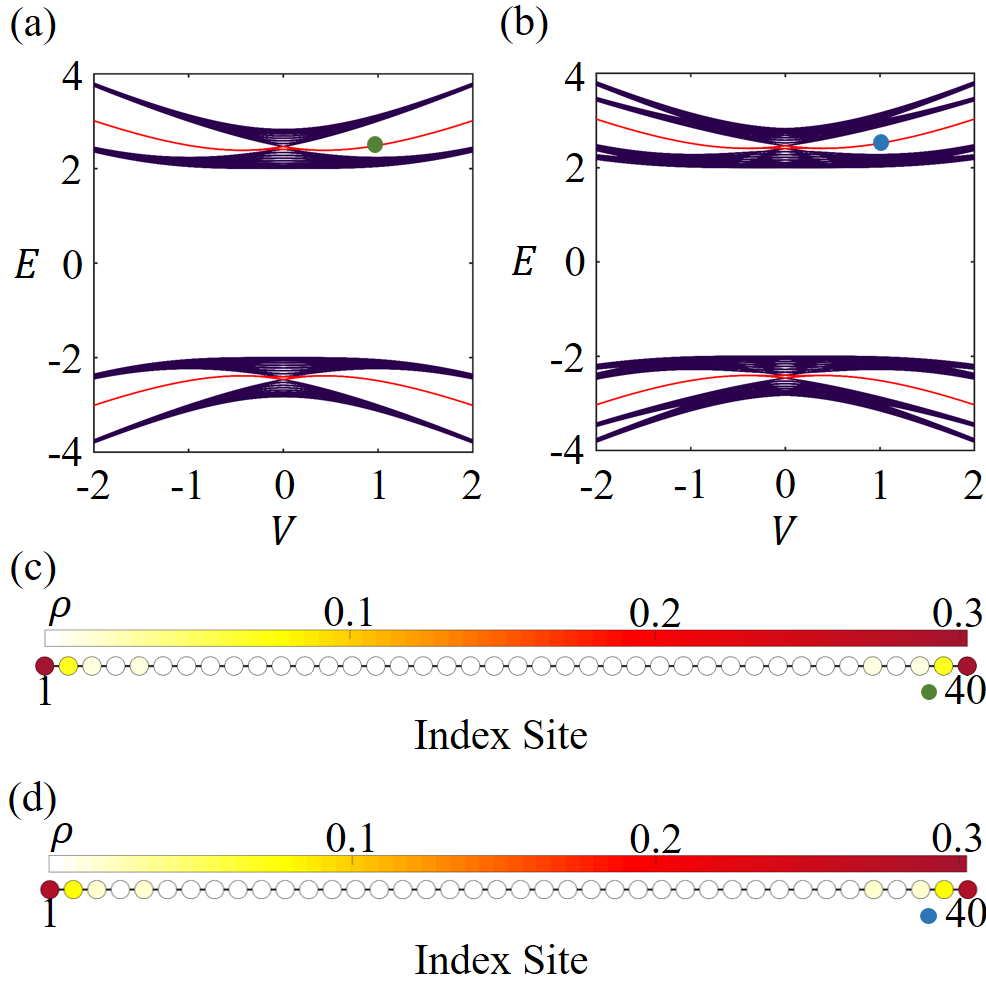}
\caption{(a) and (b) Eigenspectrum versus inversion-symmetric onsite potential $%
V$ for a 1D lattice with $40$ sites. The in-gap red lines denote four-degenerate Bogoliubov
edge excitation modes, where each edge of 1D lattice hosts two localized modes. In panel (a), we set $V_{a}=-V_{b}=V$ while in (b), we choose $V_{a}=V$, $%
V_{b}=-0.8V$. (c) and (d) The particle density distribution
versus the site index, corresponding to the colored dish in panels (a) and (b) respectively. To be specific, in panel (c) $V_{a}=-V_{b}=V=1$ and in panel (d) $V_{a}=1,$ $V_{b}=-0.8$. Common parameters
are set to be $\Delta _{0}=2$, $t=1$.}
\label{1Dedge}
\end{figure}

To demonstrate the topological properties of the system, the eigenenergies
for a chain with open boundaries are computed and plotted in Fig. \ref{1Dedge}%
(a). We observe four degenerate states emerge at the gap between the
excitation bands. Two states localize at the left end and the other two
localize at the right end of the chain, as illustrated in Fig. \ref{1Dedge}%
(c). So far, we have focused on the special cases $V_a=-V_b=V$ for simplicity. We would like to remark that if $V_{a}\neq -V_{b}$, the system also preserve the inversion symmetry, and the Bogoliubov edge states could also be driven from a the conventional $s$-wave superconductor. In Fig. \ref{1Dedge}(b) and (d), we showcase the band structure and corresponding edge states with $V_{a}\neq -V_{b}$. It demonstrates that topological edge states also emerge on the 1D lattice. Therefore, the inversion symmetry is the crucial condition to induce the Bogoliubov edge states in the $s$-wave superconductor.

We would use the Wilson loop approach to characterize the bulk topology of the
system with inversion symmetry. The base momentum point is set to be $k$. The corresponding Bloch
wave functions are denoted by $\left\vert u_{m}\left( k\right) \right\rangle
$ with $m$ representing the band index. We construct a matrix $M\left(
k\right) =\left( \left\vert u_{m}\left( k\right) \right\rangle ,\left\vert
u_{m+1}\left( k\right) \right\rangle ,...,\left\vert u_{n}\left( k\right)
\right\rangle \right) $, where $n$ stands for the number for occupied bands.  The Wilson loop operator then is defined as%
\begin{eqnarray}
\mathcal{W} &=&M\left( k\right) M\left( k+\Delta k\right) \cdots M\left(
k+\left( N-1\right) \Delta k\right)  \notag \\
&&M\left( k+N\Delta k\right) ,
\end{eqnarray}%
where $\Delta k=2\pi /N$, and $N$ is the number of unit cells. The effective
Hamiltonian is defined by $\mathcal{H}=-i\ln \mathcal{W}/\pi 
$. The eigenvalues for $\mathcal{H}$ are denoted by $v_{s} $ with $ 
s=1,2,...,n$. The bulk topological invariant is then
given by $\xi = \sum_{s=1}^{n} v_{s}$. Through numerical calculations, the topological invariant is given by $\xi=\pm2$ in superconductor phase if $V\neq0$, suggesting two localized Bogoliubov edge excitations would appear at each one edge of the 1D lattice, as demonstrated in Fig. \ref{1Dedge}.  The localized Bogoliubov edge excitations are topologically protected and
robust against inversion-symmetric perturbations as long as the bulk energy gap remains open.  


The above simple toy model exhibits interesting topological properties induced by
inversion-symmetric potentials. However, the long-range superconductor order is
forbidden due to the strong quantum fluctuations. In the following, we would
propose a realistic platform to manifest intrinsic first-order and
higher-order topology, whose topology can be explicitly understood from the
above 1D model.

\section{Bogoliubov corner excitations in an $S$-wave superfluid on a square
lattice}

\label{s2ds} 
\begin{figure}[tbp]
\centering\includegraphics[width=0.48\textwidth]{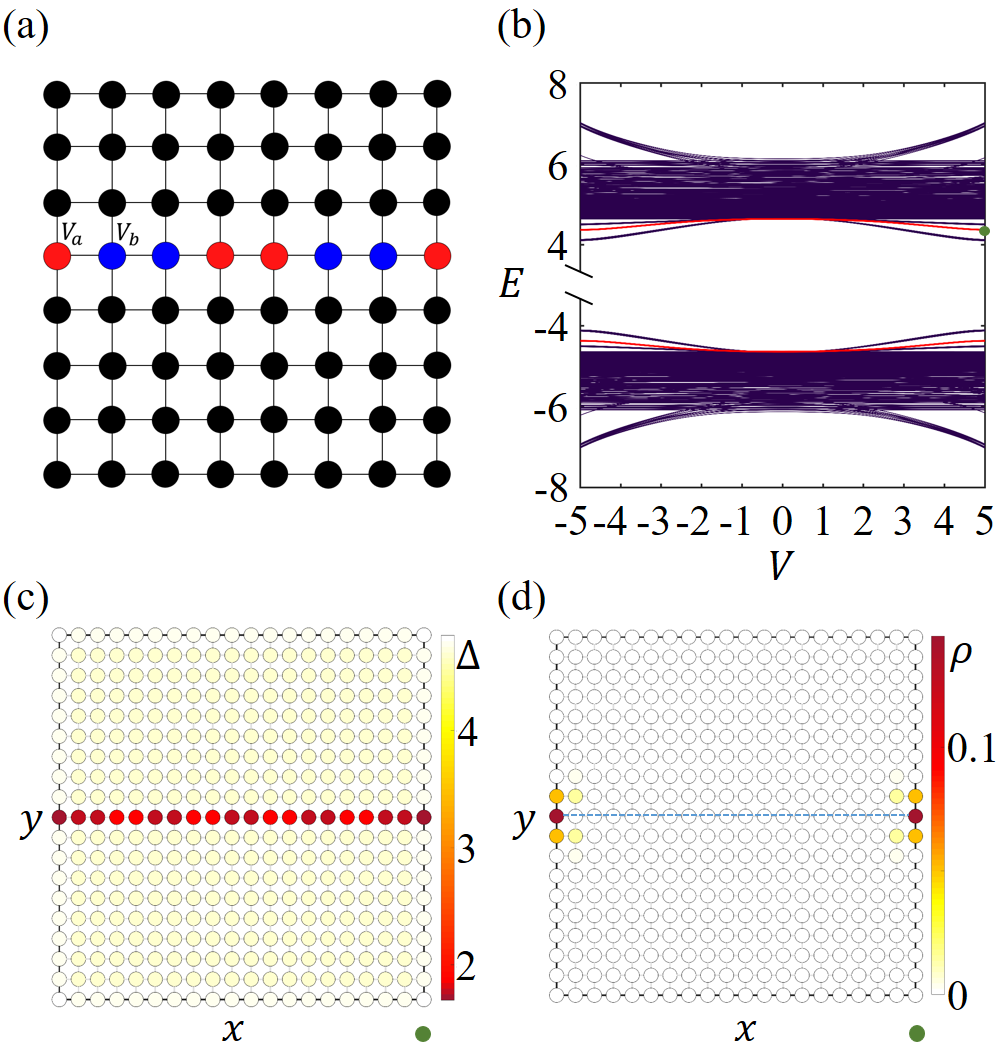}
\caption{(a) Illustration of a square lattice with a defect chain respecting
the mirror symmetry. (b) Eigenspectrum for the $s$-wave superfluid versus
mirror-symmetric onsite potential $V$ on the 20$\times $19 square lattice.
The in-gap red lines indicate four-degenerate Bogoliubov corner excitation
modes. (c) Distributions of $s$-wave superfluid order parameters
on the square lattice with a defect chain. (d) Particle density distributions of the in-gap states. The blue dashed line denotes the defect chain. The Bogoliubov excited states shown in (c) and (d) have
been indicated by the green dot in (b) with the mirror-symmetric onsite potential $V=5$. Common parameters in (b)-(d) are set
to be $t=1$, $U=10$, $\protect\mu =0.05$.}
\label{square}
\end{figure}

Here we consider ultracod Fermionic atoms with pseudo spin
loaded in a 2D square lattice. The physics for the system is described by a
tight-binding Hamiltonian as
\begin{equation}
\hat{H}_{\mathrm{squ},0}=-t\sum_{\left\langle i,j\right\rangle }\hat{c}%
_{i,\sigma }^{\dagger }\hat{c}_{j,\sigma }-U\sum_{i}\hat{n}_{i,\uparrow }%
\hat{n}_{i,\downarrow }-\mu \sum_{i,\sigma }\hat{n}_{i,\sigma },
\end{equation}%
where $U$ is the strength of an onsite attractive $\mathrm{SU(2)}$-invariant
interaction, and $\mu $ denotes the chemical potential. Given a local dip
potential with mirror symmetry applied to a one-dimensional line as shown in
Fig. \ref{square}(a), the one-dimensional defect chain also enjoys the
mirror symmetry along $x$. The total Hamiltonian then becomes
\begin{equation}
\hat{H}_{\mathrm{squ}}=\hat{H}_{\mathrm{squ},0}+\sum_{i\in \mathrm{Def}%
,\sigma }V_{i}\hat{n}_{i,\sigma },
\end{equation}%
where $V_{i}=V_{a}=V$ for $\func{mod}\left( i_{x},4\right) =1,0$\ and $%
V_{i}=V_{b}=-V$ for $\func{mod}\left( i_{x},4\right) =2,3$ on sites on the
defect chain ``$\mathrm{Def}$''. 

As the interaction $U$ becomes stronger, the fermions would be paired and
enter a superfluid phase when $U$ exceeds a critical value. The superfluid
order at the lattice site $i$ is assumed as $\Delta _{i}=U\left\langle \hat{c%
}_{i,\uparrow }^{\dagger }\hat{c}_{i,\downarrow }^{\dagger }\right\rangle $,
and the interaction term becomes $-U\sum_{i}\hat{n}_{i,\uparrow }\hat{n}%
_{i,\downarrow }=\sum_{i}\left( -\Delta _{i}^{\ast }\hat{c}_{i,\uparrow
}^{\dagger }\hat{c}_{i,\downarrow }^{\dagger }-\Delta _{i}\hat{c}%
_{i,\downarrow }\hat{c}_{i,\uparrow }+\left\vert \Delta _{i}\right\vert
^{2}/U\right) $ at the mean-field level. Through the Bogoliubov-Valatin
transformation, the creation operators $\hat{c}_{i,\uparrow }^{\dagger }$
and $\hat{c}_{i,\downarrow }^{\dagger }$ are written as $\hat{c}_{i,\uparrow
}^{\dagger }=\sum_{n=1}^{2N_{u}}\left( u_{i,\uparrow }^{n\ast }\hat{\psi}%
_{n}^{\dagger }-\upsilon _{i,\uparrow }^{n}\hat{\psi}_{n}\right) $ and $\hat{%
c}_{i,\downarrow }^{\dagger }=\sum_{n=1}^{2N_{u}}\left( u_{i,\downarrow
}^{n\ast }\hat{\psi}_{n}^{\dagger }+\upsilon _{i,\downarrow }^{n}\hat{\psi}%
_{n}\right) $, where $N_{u}$ is the number of unit-cells, and $\hat{\psi}%
_{n}^{\dagger }$ and $\hat{\psi}_{n}$ are creation and annihilation
operators for Bogoliubov quasi-particles such that the Hamiltonian $\hat{H}_{%
\mathrm{squ}}$ can be diagonalized. The coefficients $u_{i,\sigma }^{n}$ and
$\upsilon _{i,\sigma }^{n}$ can be derived from the following equations%
\begin{eqnarray}
\sum_{j}\hat{H}_{\mathrm{0},ij,\sigma }u_{j,\sigma }^{n}-\Delta _{i}\upsilon
_{i,\bar{\sigma}}^{n} &=&E_{n}u_{i,\sigma }^{n}, \\
-\sum_{j}\hat{H}_{\mathrm{0},ji,\sigma }\upsilon _{j,\sigma }^{n}-\Delta
_{i}^{\ast }u_{i,\bar{\sigma}}^{n} &=&E_{n}\upsilon _{i,\sigma }^{n},
\end{eqnarray}%
where $\hat{H}_{\mathrm{0},ij,\sigma }$ denotes the element of the
Hamiltonian matrix $\hat{H}_{\mathrm{squ}}$ with $U=0$ under the basis $\hat{%
\Psi}=\left( \hat{C}_{1},...,\hat{C}_{m},...,\hat{C}%
_{2N_{u}}\right)^{T} $ with $\hat{C}_{m}=\left( \hat{c}_{m,\uparrow },\hat{c}%
_{m,\downarrow }\right) $.

Through the numeric calculations, we compute the superfluid order parameter
at each lattice site on a square optical lattice under open boundary
conditions, as shown in Fig. \ref{square}(c). The superfluid order on the
defect chain is weaker than that in other regions due to the non-zero mirror
symmetric potentials. In addition, we observe that the superfluid order on the
boundary is stronger than that in the bulk, and the superfluid order in the bulk is
nearly uniform. This is consistent to the intuition that the lattice sites in the bulk are less affected by the boundary. The eigen-energy distributions versus potential $V$ have been shown
in Fig. \ref{square}(b), indicating isolated states (denoted by red lines) emerge in the energy gap
for Bogoliubov quasiparticles. The particle density distributions for the
isolated states, as plotted in Fig. \ref{square}(d), showcase these in-gap
states are localized at the end of the defect chain.

We would like to remark that the above defect chain can be considered as a
one-dimensional $s$-wave superfluid imprinted on the 2D lattice. While the
defect chain couples with other chains, it also exhibits topological
nontrivial properties as long as the energy gap remains open.

\section{Bogoliubov corner excitations in an $S$-wave superfluid on a
honeycomb lattice}

\label{s2h}
\begin{figure}[tbp]
\centering\includegraphics[width=0.48\textwidth]{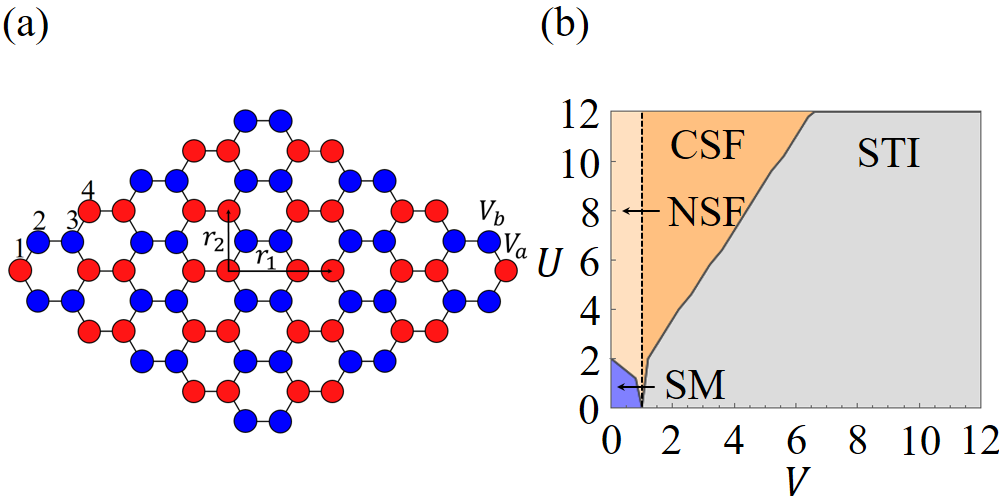}
\caption{(a) Illustration of a honeycomb lattice with a mirror-symmetric
onsite potential. Each unit-cell consists of four sublattice sites indexed
by 1 to 4. (b) A rich global phase diagram plotted against the potential and interaction strength including second-order topological
insulators (STI), semimetal (SM), normal superfluid (NSF) and superfluid
with Bogoliubov corner excitations (CSF) .}
\label{honeycomb}
\end{figure}

Consider a two-component Fermi gas loaded in a 2D honeycomb optical lattice
with a uniform chemical potential. Turning on the onsite attractive
interaction for fermionic atoms, the fermions would be paired and enter the $%
s$-wave superfluid phase from the semimetal phase when the interaction
exceeds a critical value \cite{Zhao2006}. Its Bogliubov excitations are
gapless and show trivial properties. Hereafter, we would consider there
exists onsite potential with mirror symmetry as shown in Fig. \ref{honeycomb}%
(a), and showcase Bogoliubov corner excitations could be induced from the excitation bands.

At the mean-field level, the physics of a system with a mirror-symmetric potential is described by the Hamiltonian as

\begin{eqnarray}
h_{h} &=&\lambda _{0}(s_{y}s_{x}\sigma _{0}\tau _{z}+s_{x}s_{y}\sigma
_{0}\tau _{z})+A_{k}+\lambda _{a}s_{x}s_{x}\sigma _{0}\tau _{z}  \notag \\
&&+\lambda _{b}s_{x}s_{0}\sigma _{0}\tau _{z}+\lambda _{c}s_{y}s_{0}\sigma
_{0}\tau _{z}+\lambda _{k}s_{y}s_{y}\sigma _{0}\tau _{z}
\end{eqnarray}
under the basis vector $\hat{C}_{\mathbf{k}}^{\dagger }=\left( \hat{\psi}_{%
\mathbf{k}\uparrow }^{\dag },\hat{\psi}_{\mathbf{k}\downarrow }^{\dagger },%
\hat{\psi}_{-\mathbf{k}\uparrow },\hat{\psi}_{-\mathbf{k}\downarrow }\right)
$ with $\hat{\psi}_{\mathbf{k}\sigma }^{\dag }=$ $\left( \hat{c}_{1,\mathbf{k%
},\sigma }^{\dagger },\hat{c}_{2,\mathbf{k},\sigma }^{\dagger },\hat{c}_{3,%
\mathbf{k},\sigma }^{\dagger },\hat{c}_{4,\mathbf{k},\sigma }^{\dagger
}\right) $ and $\hat{\psi}_{-\mathbf{k}\sigma }=$ $\left( \hat{c}_{1,-%
\mathbf{k},\sigma },\hat{c}_{2,-\mathbf{k},\sigma },\hat{c}_{3,-\mathbf{k}%
,\sigma },\hat{c}_{4,-\mathbf{k},\sigma }\right) $, where $\lambda
_{0}=t\sin k_{\alpha }$, $\lambda _{a}=\frac{1}{2}t(1+\cos k_{\alpha })$, $%
\lambda _{b}=t(1+\cos \sqrt{3}k_{y})$, $\lambda _{c}=t\sin \sqrt{3}k_{x}$, $%
\lambda _{k}=\frac{1}{2}t(1-\cos k_{\alpha })$, $k_{\alpha }=\mathbf{k}\cdot
\left( \mathbf{r}_{1}+\mathbf{r}_{2}\right) =3k_{x}+\sqrt{3}k_{y}$, $%
A_{k}=\Delta _{1}s_{+}s_{+}\sigma _{y}\tau _{z}+\Delta _{2}s_{-}s_{+}\sigma
_{y}\tau _{z}+\Delta _{3}s_{+}s_{-}\sigma _{y}\tau _{z}+\Delta
_{4}s_{-}s_{-}\sigma _{y}\tau _{z}+V_{+}s_{0}s_{0}\sigma _{0}\tau
_{z}+V_{-}s_{z}s_{z}\sigma _{0}\tau _{z}$ with $V_{\pm }=\left( V_{a}\pm
V_{b}\right) /2$ and $s_{\pm }=\left( s_{0}\pm s_{z}\right) /2$. $\Delta
_{m=1,2,3,4}$ denotes the superfluid order parameter on the sublattice site
$m$ as indexed in Fig. \ref{honeycomb}(a). The self-consistent equations
for the superfluid order and particle filling ratio are given by
\begin{eqnarray}
\Delta _{m} &=&\frac{U}{N_{u}}\sum_{k\in BZ/2}\left\langle \hat{c}_{m,-%
\mathbf{k},\downarrow }\hat{c}_{m,\mathbf{k},\uparrow }\right\rangle ,
\label{self1} \\
n_{m} &=&\frac{1}{N_{u}}\sum_{k\in BZ/2}\left\langle \hat{c}_{m,\mathbf{k}%
,\uparrow }^{\dagger }\hat{c}_{m,\mathbf{k},\uparrow }+\hat{c}_{m,\mathbf{k}%
,\downarrow }^{\dagger }\hat{c}_{m,\mathbf{k},\downarrow }\right\rangle .
\label{self2}
\end{eqnarray}%
Through numerical self-consistent calculations for Eqs. (\ref{self1}) and (%
\ref{self2}), we find $\Delta _{m}\equiv
\Delta $ and $n_{m}\equiv n$ for all $m$ at zero temperature. The rich
phase diagram, which has been shown in Fig. \ref{honeycomb}(b), exhibits a range of interesting and physically distinctive phases including semimetal (SM), second-order topological insulator (STI), and superfluid
phases with a non-zero superfluid order (normal superfluid and superfluid
with Bogoliubov corner excitations). Fig. \ref{honeycomb}(b) showcases that at fixed interaction, for example $U=4$, as $V$ becomes larger, the system would enter STI since it's hard to form pairing when $V$ exceeds a critical value.

\begin{figure}[tbp]
\centering\includegraphics[width=0.48\textwidth]{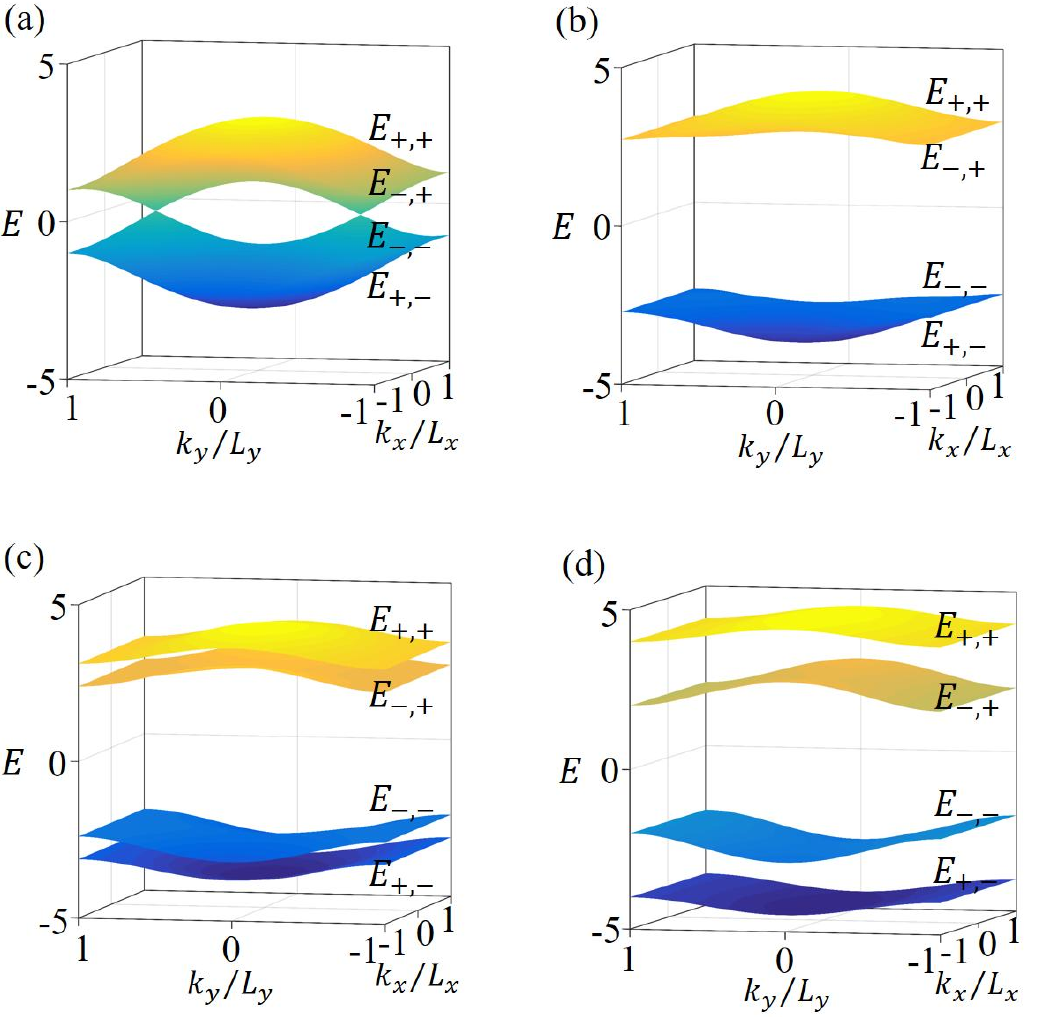}
\caption{Energy spectra for different onsite potentials $V$. In (a), $U=V=0$%
. In (b), $U=6$, $V=0$. In (c), $U=6$, $V=1$. In (d), $U=6$, $V=3$. Common
parameters are set to be $t=1$, $L_{x}=\protect\pi /3$, $L_{y}=\protect\pi /%
\protect\sqrt{3}$.}
\label{phased}
\end{figure}

We compute the energy spectra for the superfluid under periodic boundary
conditions and the numerical results are presented in Fig. \ref{phased}. We observe the $s$%
-wave superfluid order opens the energy gap for the Dirac semimetal, as
shown in Figs. \ref{phased}(a) and (b). However, the Bogoliubov excitation
band remains gapless. After turning on the onsite potential with mirror
symmetry, an direct energy gap emerges, as depicted in Fig. \ref{phased}(c).
As the potential strength increases, the energy gap becomes larger and a
full gap exists when the potential exceeds a critical value, as illustrated
in Fig. \ref{phased}(d).

\begin{figure}[tbp]
\centering\includegraphics[width=0.48\textwidth]{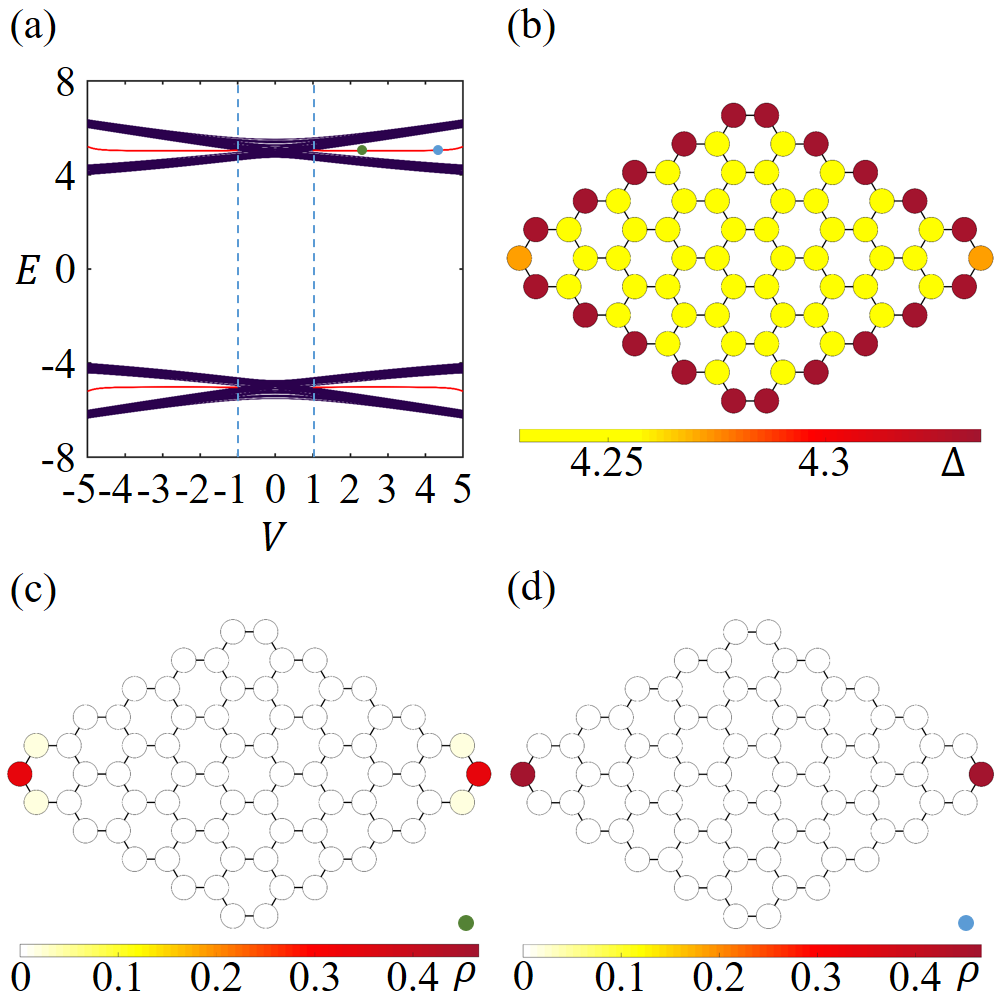}
\caption{(a) Eigenspectrum versus mirror-symmetric onsite potential $V$ for
the honeycomb lattice. The in-gap red lines denote four-degenerate
Bogoliubov corner excitation modes. (b) Distributions of $s$-wave
superfluid order parameters on the honeycomb lattice. (c) and (d) Particle
density distributions of the in-gap states. These chosen parameters in
sub-figures also have been indicated by colored dots in (a). In (b) and (c),
$V_{a}=-V_{b}=V=2.2$. In (d), $V_{a}=-V_{b}=V=4.5$. Common parameters are
set to be $t=1$, $U=10$.}
\label{hspec}
\end{figure}

To explore the nontrivial properties of Bogoliubov excitation bands, we
calculate the eigenenergies for the superfluid versus $V$ with fixed
interaction $U$ under open boundary conditions. Four degenerate states
emerge in the energy gap for Bogolibov excitations, as shown by the red
lines with four-fold degenerates in Fig. \ref{hspec}(a). They are localized
at two corners of the sample as shown in Fig. \ref{hspec}(c). Through the
numeric calculations, we compute the superfluid order parameters at each
lattice site on a honeycomb optical lattice under open boundary conditions,
as shown in Fig. \ref{hspec}(b), we can observe that the bulk superfluid
order is uniform. Through comparing Fig. \ref{hspec} (c) with (d), it's also clear that with increasing potential $V$, the Bogoliubov corner
modes become more localized. In
summary, there are superfluid phases with two different Bogoliubov
excitations, one with gapless Bogliubov excitation bands dubbed NSF, and the
other with gapped Bogliubov excitation bands called CSF. Both phases and the their phase transition are demonstrated in the
phase diagram in Fig. \ref{honeycomb}(b).

The Bogoliubov corner excitations is the exhibition of the bulk topology,
characterized by the topological invariant protected by mirror symmetry.
Taking a similar procedure as in the one-dimensional case above, the topological
invariant at each $k_{y}$ is defined by
\begin{equation}
\xi _{x}\left( k_{y}\right) =-\frac{i}{\pi }\mathrm{Tr}\left( \ln \mathcal{W}%
_{x,k}\right) ,
\end{equation}
where the Wilson loop operator reads $\mathcal{W}_{x,k}=F_{x,k+N_{x}\Delta
k_{x}}...F_{x,k+\Delta k_{x}}F_{x,k}$, $\Delta k_{x}=2\pi /N_{x}$, and
$N_{x}$ is the number of unit cells in the $x$ direction. The entry of
matrix $F_{x,k}$ is $\left[ F_{x,k}\right] ^{m,n}=\left\langle u_{m,k+\Delta
k_{x}}|u_{m,k}\right\rangle $ with $\left\vert u_{m,k}\right\rangle$ being the
Bloch wave function of the energy bands $E_{m}\left( k\right) $, i.e., $%
h_{h}\left( k\right) u_{m,k}=E_{m}\left( k\right) $ while $\xi _{x}\left(
k_{y}\right) $ forms the Wannier bands. Finally, the topological invariant
is defined as $(\xi _{x}^{\prime },\xi _{y}^{\prime })$, where $\xi
_{x}^{\prime }=\frac{1}{2N_{y}}\sum_{k_{y}}\xi _{x}\left(
k_{y}\right) $, and $\xi _{y}^{\prime }$ takes similar form as $%
\xi _{x}^{\prime }$. Through numeric calculations, we obtain the topological
invariant $(\xi _{x}^{\prime },\xi _{y}^{\prime })=(2,0)$ in the CSF
regime and $(0,0)$ in NSF regime, as shown in Fig. \ref{honeycomb}(b).
In summary, the $s$-wave superfluid phase have two different types of Bogoliubov excitations: trivial Bologliubov excitations in NSF regime, and higher-order Bogoliubov corner excitations in CSF regime. We emphasize that the ground state of the $s$-wave superfluid in both regimes is topologically trivial. The topological property of excited corner modes originates from the Bogoliubov excitation bands.

\section{Discussions and Conclusions}

\label{con} The $s$-wave superfluid with a uniform chemical potential
exhibits trivial Bogoliubov excitation on a 1D lattice, and 2D square or
honeycomb lattice. Intriguingly, we find the onsite potential with mirror
symmetry could open the energy gap in Bogoliubov excitation spectrums. The Bogliubov
excitation bands exhibit topological nontrivial properties and the edge modes
manifest themselves as zero-dimensional (0D) Bogliubov excitations localized at the end of a 1D
lattice and the corners of a 2D honeycomb lattice, although the ground states for the systems remains in a trivial phase. Since the systems preserve
inversion or mirror symmetry, the winding number can characterize the nontrivial
excitation band.

We would like to remark that our model in this work can be implemented in
ultracold atoms. For instance, the mirror-symmetric potential on 2D square
lattice can be achieved through a pair of coherent counterpropagating laser
beams with wave length $2a$ and $8a$ along $x$. The onsite attractive
interaction could be finely tuned through Feshbach Resonance technique.

In summary, we propose that topological Bogoliubov excitations can be induced
solely by onsite potentials in a topologically trivial conventional $s$-wave
superfluid. The edge excitations manifest themselves as 0D modes localized
at edges or corners of the system. These modes are robust against inversion or mirror symmetric
perturbations as it preserves the degeneracy.
Our work provides new insights for understanding higher-order
topological states in conventional superconductors and superfluids, and also provides realistic platforms for engineering nontrivial Bogoliubov corner excitations in real experiments.

\begin{acknowledgments}
This work was supported by NSFC under the grant Nos. 12275203 and 12075176,
the Scientific Research Program Funded by Natural Science Basic Research
Plan in Shaanxi Province of China (Program No. 2021JM-421), Innovation
Capability Support Program of Shaanxi (2022KJXX-42), and 2022 Shaanxi
University Youth Innovation Team Project (K20220186).
\end{acknowledgments}

\end{document}